\documentstyle[12pt]{article}
\begin{document}
\author{S. M. Troshin, N. E. Tyurin\\
Institute for High Energy Physics,\\
142284 Protvino, Moscow Region, Russia }
\title
{Beyond the black disk limit: antishadow scattering mode}
\maketitle
\begin{abstract}

The appearance of the antishadow scattering mode
at energies of the Tevatron--Collider is considered on the basis
of unitarity and geometrical notions on hadron interactions.
\end{abstract}

\section*{Introduction}

One of the most fundamental discoveries in hadron interactions
at high energies is the rise of total cross--sections with energy.
It is accompanied  by the rise of elastic and inelastic cross--sections
as well as of the ratio of elastic to total cross--sections.

For the first time the total cross--section rise was observed in
$K^+p$--interactions at the Serpukhov accelerator in 1970 and was
discovered later also in  $pp$--interactions at CERN ISR and at Fermilab
in other hadronic reactions.

Since that time a great progress in the experimental and theoretical
studies of soft hadronic reactions has been achieved. Quantum
Chromodynamics has appeared as a theory of strong interactions.
However, dynamics of long distance hadronic interactions is rather
far from its complete understanding and much work is needed in
this field. The problems here are  directly
connected with the problems of confinement and chiral symmetry
 breaking.

Despite of inaplicability of perturbative QCD for the description
of long--distance interactions and their obvious nonperturbative character,
 it is often possible  to
represent  the high--energy amplitude in the model
approaches as an expansion in some parameter
which depends on the kinematics of the process, e.g. for the case of
non--increasing total cross--section the general form of amplitude
is
\[
F(s,t)=s\sum_n [\tau (s)]^n\exp \left[\frac{a(s)t}{n}\right],
\]
where $\tau(s)\sim 1/\ln s $ is a small parameter at $s\rightarrow \infty$.

This expansion is not valid for the rising total cross--sections.
However, it is possible to find another representation in that case
with $t$--dependent expansion parameter \cite{expt}:
\[
F(s,t)=s\sum_{m=1}^\infty [\tau (\sqrt{-t})]^m\Phi _m[R(s), \sqrt{-t}],
\qquad t\neq 0,
\]
where
\[
\tau (\sqrt {-t})=\exp\left(-\frac{\sqrt{-t}}{\mu _0}\right).
\]
and $\Phi _m[R(s), \sqrt{-t}]$ is an oscillating function of transferred
momentum.
The both above formulas as well as some other representations may be
succesfully
used for the phenomenological analysis of the scattering amplitude
at high energies.

Thus, by now the theoretical treatment of soft hadronic reactions
involves substantial piece of phenomenology and uses various model
 approaches such as Regge--type, geometrical or QCD--inspired models.
They are based on divergent postulates, but their phenomenological
parts are similar. In particular, some amplitude $V(s,t)$ is considered
as an input for the subsequent unitarization procedure:
\[
F(s,t)=\Phi [V(s,t)].
\]
To reproduce the total cross--section growth the input amplitude
$V(s,t)$ is usually considered as  some power--behaved function of energy.
This function taken as an amplitude itself violates
 unitarity  in the direct channel.
To obey unitarity in the direct channel the unitarization procedures are
to be used.

There are several ways to restore unitarity  of the scattering
matrix. We are going to consider the  two such schemes: based on the
 use of eikonal and
the method of generalized reaction matrix respectively.
As it was already mentioned various models for $V(s,t)$
 may be successfully used to provide
 phenomenological description of high energy hadron scattering.
However, in the particular model
approaches the important dynamical aspects of interaction
are obscured often due to  large number of free parameters.

In this brief review we discuss some general
properties of hadron scattering, the implications
 of unitarity and analyticity,
 in particular, manifestations of the antishadow
 scattering mode (Section 3).
The recent CDF data indicate that the black disk limit
 is probably already violated at the Tevatron--Collider energy
$\sqrt{s}=1.8$ $TeV$.
Preliminary discussions of  geometrical picture and the bounds
for observables are given in Sections 1 and 2 to define a frame
of the problem.

\section{Geometrical Picture}

In the collisions of two high energy particles the de Broglie
wavelenth can be short compared to the typical hadronic size
and hence some optical concepts may be used as  useful guidelines.
We can consider therefore hadron scattering
as a collision of two
relativistically contracted objects of finite size.

The relevant mathematical tool for the description of high energy
hadronic scattering is based on the
impact parameter representation for the scattering amplitude.
This representation has the following form in the spinless case:
\begin{equation}
F(s,t)=\frac{s}{\pi ^2}\int _0^\infty bdb f(s,b)J_0(b\sqrt{-t}).
\label{imp}
\end{equation}
Note that for the scattering of particles with non--zero spin
the impact parameter representation for the helicity amplitudes
 has similar form with substitution $J_0\rightarrow J_{\Delta\lambda}$,
where $\Delta\lambda$ is the net helicity change between the final
and initial states.
The impact parameter representation as it was shown in \cite{islm}
is valid for all physical energies and scattering angles. This representation
 provides simple semiclassical picture of hadron scattering.

It is often assumed, since the Chou--Yang model was proposed, that the
main effect in hadron scattering arises due to overlapping of the two
 matter distributions. It could be understood by analogy with the Glauber
 theory of nuclear interactions: one assumes that the matter density
comes from the spatial distribution of hadron constituents and
assumes also a zero--range interaction between the constituents.
 Such contact interaction might result from the effective
QCD based for example on the Nambu--Jona-Lasinio model.

The general definition of the interaction radius which is in
 agreement with the above geometrical picture was given in \cite{ngu}:
\begin{equation}
R(s)=l_0(s)/k,
\end{equation}
where $k=\sqrt{s}/2$ is the particle momentum in the c.m.s.
The value for $l_0(s)$ is chosen provided the contribution to the
partial amplitude from the angular momenta $l> l_0(s)$ are
vanishinly small.

 In the first approximation one can consider the
energy independent intensity and describe the elastic scattering amplitude
in terms of the black disk model where it has
  the form:
\begin{equation}
F(s,t)\propto iR^2(s)\frac{J_1(R(s)\sqrt{-t})}{R(s)\sqrt{-t}}.
\end{equation}
Here $R\sim 1 f$ is the interaction radius. The model is consistent
with the structure in the differential cross--sections of $pp$-- and
$\bar p p$--scattering observed at $t$ near  1 $(GeV/c)^2$.

When neglecting the real part and helicity flip amplitudes the impact
parameter amplitude $f(s,b)$ can be obtained as an inverse
transformation according to Eq. \ref{imp} with
\[
F(s,t)\propto\sqrt{s\frac{d\sigma}{dt}(s,t)}.
\]
Thus, one can extract information on the geometrical properties
of interaction from the experimental data.
The analysis of the experimental data on high--energy diffractive scattering
 shows that the effective interaction area expands with energy and the
interaction intensity --- opacity --- increases with energy at
fixed impact parameter $b$. Such analysis used to be carried out
every time as the new experimental data become available. For example
analysis of the  data at the ISR energies (the most
precise data set on differential cross--section for wide $t$--range
available for $\sqrt{s}=53$ GeV) shows that one can observe a central
 impact parameter profile with a tail from the higher partial waves and some
suppression of low partial waves relative to a gaussian.
 The scattering picture at such energies is close to   grey
 disk with smooth edge which is getting  darker in the center with
energy.

After these simple geometrical observations we consider the
 bounds for the experimental observables.

\section{Bounds for observables and the experimental data}

Bounds for the observables obtained on the firm ground of
general principles such as unitarity and analiticity
are very important for any phenomenological analysis
of soft interactions. However, there are only  few results
obtained on the basis of the axiomatic field theory.

First of all it is the Froissart--Martin bound that gives the
 upper limit
 for the total cross--section:
\begin{equation}
\sigma _{tot}\leq C\ln^2 s,
\end{equation}
where $C=\pi/m_{\pi}^2$ $(=60$mb) and $m_{\pi}$ is the pion mass.

Saturation of this bound, as it is suggested by the existing
 experimental data, imply the dominance of long--distance dynamics.
It also leads to  number of important consequences for the other
observables. For instance, unitarity leads to the following bound
 for elastic cross--section:
\begin{equation}
\sigma_{el}(s)\geq c \frac{\sigma _{tot}^2(s)}{\ln^2 s}.
\end{equation}

Therefore, when the total cross--section increases as $\ln^2 s$,
elastic cross--section also must rise like $\ln^2 s$. It is
important to note here that there is no similar restriction for the
inelastic
cross--section and as we will see further the absence of such bound
 allows appearance of the antishadow scattering mode at very high
 energies.

If one considers a more general case when
 $\sigma_{tot}\propto \ln ^\gamma s$, then
at asymptotic energies one should have
\begin{equation}
\frac{Re F(s,0)}{Im F(s,0)}\simeq \frac{\gamma\pi}{2\ln s}
\end{equation}
and
\begin{equation}
\frac{\sigma^{\bar a}_{tot}(s)-\sigma^{a}_{tot}(s)}
{\sigma^{\bar a}_{tot}(s)+\sigma^{a}_{tot}(s)}\leq
\ln^{-\gamma/2}(s)
\end{equation}
where $\sigma^{\bar a}_{tot}(s)$ and $\sigma^{a}_{tot}(s)$ are the
total cross--sections of the processes $\bar a +b\rightarrow X$ and
 $a +b\rightarrow X$ correspondingly. In the case of $\gamma =2$ the
total cross--section difference of antiparticle and particle interaction
 should obey the following inequality
\begin{equation}
\Delta\sigma_{tot}(s)\leq\ln s.
\end{equation}

Contrary to the total cross--section behavior the existing experimental data
seem to prefer decreasing $\Delta\sigma_{tot}(s)$. Possible deviations
from such behavior could be expected on the basis
of perturbative QCD \cite{lipa}
 and it was one of the reasons for the recent discussions
 on the Pomeron counterpart --- the Odderon. However, the recent measurements
of real to imaginary part ratio for forward $\bar p p$ scattering
 provide little support for the Odderon. We will not discuss
more thoroughly $Re F/Im F$ ratio and  will  consider
for simplicity the case of pure imaginary amplitude.

For  pure imaginary scattering amplitude the following inequality
takes place for the slope of diffraction cone at $t=0$:
\begin{equation}
B(s)\geq\frac{\sigma^2_{tot}(s)}{18\pi\sigma_{el}(s)}
\end{equation}
which means that when the total cross--section increases
as $\ln^2 s$, the same dependence is obligatory for the
slope of diffraction cone. It is stronger shrinkage than
the Regge model predicts $B(s)\sim \alpha '\ln s$.

 There is also bound \cite{pmpl} for the total cross--section
of single diffractive processes. It was obtained in  approach
where inelastic diffraction as well as elastic scattering are assumed
to arise as a shadow of inelastic processes and has the
form
\begin{equation}
\sigma_{diff}(s,b)\leq\frac{1}{2}\sigma_{tot}(s,b)-\sigma_{el}(s,b).
\label{pmb}
\end{equation}
In particular, it was assumed that the diffractive eigenamplitudes
in the Good--Walker \cite{gwlk} picture do not exceed the black disk
limit.

At this point some details of the experimental situation are to be
mentioned. Recently the new experimental data for the total and elastic
 cross--sections, slope parameter of diffraction cone and cross--section of
 single inelastic diffraction
dissociation have been collected in $\bar p p$--collisions at
Fermilab. We will refer mainly to the recent CDF results for diffractive
scattering. In particular these measurements show that

\begin{itemize}
\item
total cross--section of $p\bar p$--interactions is large
$\sigma _{tot}=80.6\pm 2.3$ mb at $\sqrt{s}=1.8$ TeV which
is consistent with $\ln^2 s$--rise;
\item
elastic cross--section also has a large value:
$\sigma_{el}=20.0\pm 0.9$ mb and ratio of elastic to total
cross--section $\sigma _{el}/\sigma_{tot}=0.248\pm 0.005$;
\item
the impact parameter scattering amplitude $Im f(s,b=0)=0.50\pm0.01$.
\end{itemize}
Comparing these values with the lower energy data one can conclude
also that the higher the energy, the larger both absolute and
relative probabilities of elastic collisions.

Impact parameter analysis of the data shows that the scattering amplitude
is probably beyond the black disk limit $|f(s,b)|=1/2$ in head-on collisions.
The Pumplin bound Eq. \ref{pmb} is also violated in such collisions and
this is not surprusing if one recollects the original assumption on the
shadow scattering mode.

 It should be noted that another experiment at Fermilab E710 \cite{roy}
 gives different values for the cross--sections and therefore the above
 conclusions should be taken with certain precautions.

However, it seems worth to consider the possibility that the
 scattering amplitude exceeds the black disk limit in head--on
 collisions and that the transition to the antishadow scattering
mode might occur in the central hadron collisions.

\section{Antishadow scattering mode}

The basic role in our consideration belongs to unitarity
of the scattering matrix $SS^+=1$
which reflects the probability conservation.
In the impact parameter representation Eq. \ref{imp}
unitarity  has a simple form
\begin{equation}
Im f(s,b)=|f(s,b)|^2+\eta(s,b) \label{unt}
\end{equation}
where the inelastic overlap function $\eta(s,b)$ is the sum of
all inelastic channel contributions.  It can be expressed as
a sum of $n$--particle production cross--sections at given impact
 parameter
\begin{equation}
\eta(s,b)=\sum_n\sigma_n(s,b).
\end{equation}
As it was already mentioned  consideration of pure
 imaginary amplitude is rather good approximation at high energies.
Then the unitarity Eq. \ref{unt} points out that the elastic scattering
 amplitude at given impact parameter value
is determined by the inelastic processes.
Eq. \ref{unt} imply the constraint
\[
|f(s,b)|\leq 1
\]
 while the black disk limit
 presumes inequality
\[
|f(s,b)|\leq 1/2.
\]
 The equaility $|f(s,b)|=1/2$ corresponds
to maximal absorbtion in the partial wave with angular momentum
$l\simeq b\sqrt{s}/2$.

The maximal absorbtion limit is chosen a piori in the eikonal method
of unitarization when the scattering amplitude is written in the
 form:
\begin{equation}
f(s,b)=\frac{i}{2}(1-\exp[i\omega(s,b)])
\end{equation}
and imaginary eikonal $\omega(s,b)=i\Omega(s,b)$ is considered.
The function $\Omega(s,b)$ is called opacity. Eikonal unitarization
 automatically satisfies the unitarity  Eq. \ref{unt} and in
the case of pure imaginary eikonal leads to amplitude which is always
under the black disk limit.

However, unitarity equation has
two solutions for  pure imaginary case:
\begin{equation}
f(s,b)=\frac{i}{2}[1\pm \sqrt{1-4\eta(s,b)}].\label{usol}
\end{equation}
Eikonal unitarization with pure imaginary eikonal corresponds to the
choice of the particular
solution with sign minus.

Several models have been proposed for the
eikonal function. For instance, Regge--type models lead to the gaussian
dependence of $\Omega(s,b)$ on impact parameter.  To provide rising
total cross--sections opacity should have a power dependence on energy
\begin{equation}
\Omega(s,b)\propto s^\Delta\exp[-b^2/a(s)],
\end{equation}
where $a(s)\sim \ln s$. In the framework of perturbative QCD--based models
the driving contribution to the opacity is due to jet production in
 gluon--gluon interactions, when
\begin{equation}
\Omega(s,b)\propto \sigma_{jet}\exp[-\mu b],
\end{equation}
where $\sigma_{jet}\sim (s/s_0)^\Delta$.
This parameterization leads to the rising total and elastic
 cross--sections and slope parameter:
\begin{equation}
\sigma_{tot}(s)\sim \sigma_{el}(s)\sim B(s)\sim \ln^2 s
\end{equation}
and the ratio
\begin{equation}
\frac{\sigma_{el}(s)}{\sigma_{tot}(s)}\rightarrow \frac{1}{2}.
\end{equation}

 Transition to the mode where the scattering amplitude exceeds the
 black disk limit results in the necessity of considerations of the
eikonal functions with non--zero
 real parts. Then to ensure such transition the real part
of eikonal should gain an abrupt increase
equal to $\pi$ at some $s=s_0$. The conventional models do
not foresee such a critical behavior of the real part of eikonal.

However, it does not
 mean that the eikonal model itself is in trouble. In particular, the
account for the fluctuations of the eikonal \cite{bars} strongly
 modifies the structure
 of the amplitude and reduces it to algebraic form which is similar to
 that used in another unitarization scheme with the use of the
 generalized reaction
matrix.

This method
 is based on the relativistic generalization
 of the Heitler equation \cite{logn}.
The form of the amplitude in the framework
 of this method
 is the following:
\begin{equation}
f(s,b)=\frac{U(s,b)}{1-iU(s,b)} \label{um}
\end{equation}
where $U(s,b)$ is the generalized reaction matrix, which is considered as an
input dynamical quantity similar to an eikonal function.
Inelastic overlap function
is connected with $U(s,b)$ by the relation
\begin{equation}
\eta(s,b)=Im U(s,b)|1-iU(s,b)|^{-2}.
\end{equation}

Eq. \ref{um} ensures $s$--channel unitarity provided that
$Im U(s,b)\geq 0$.
 Similar form for the scattering matrix was obtained by Feynman in his
 parton model of diffractive scattering \cite{ravn}.

 Construction of  particular models in the framework of the $U$--matrix
approach proceeds with  the same steps as it happens
for the eikonal function, i.e. the basic dynamics as well as
the notions on hadron structure
are used  to obtain a particular form for the $U$--matrix.
For example, the Regge--pole approach
\cite{tukh} provides the following form for the $U$--matrix:
\begin{equation}
U(s,b)\propto is^\Delta\exp [-b^2/a(s)],\quad a(s)\sim \alpha '\ln s,
\label{eb2}
\end{equation}
while the chiral quark model \cite{chqm} gives the exponential
$b$--dependence
\begin{equation}
U(s,b)\propto is^\Delta\exp [-\mu b],\label{eb1}
\end{equation}
where $\mu$ is the constant proportional to the mass of the constituent
 quarks.  We pointed out here only the gross features of these model
 parameterizations without going into details.

The both parameterizations lead to $\ln^2 s$ rise of the total and
elastic cross--sections and slope parameter $B(s)$:
\begin{equation}
\sigma_{tot}(s)\sim\sigma_{el}(s)\sim B(s) \sim \ln^2 s
\end{equation}
at $s\rightarrow\infty$.
These results are similar to the results of eikonal unitarization.

However, these two unitarization schemes give different predictions for
 the inelastic cross--sections and for the ratio of elastic to total
cross-section. This ratio in the $U$--matrix unitarization scheme
reaches its maximal possible value at $s\rightarrow \infty$, i.e.
\begin{equation}
\frac{\sigma_{el}(s)}{\sigma_{tot}(s)}\rightarrow 1,
\end{equation}
which reflects in fact that the bound for the partial--wave
 amplitude in the $U$--matrix
approach is $|f|\leq 1$
while  the bound for the case of imaginary eikonal
is (black disk limit):
$|f|\leq 1/2$.

When the amplitude exceeds the black disk limit (in central
collisions at high energies) then the scattering at such impact
parameters turns out to have antishadow nature. It means that we
 should consider in this case the solution of unitarity equation
Eq. \ref{unt} which has sign plus. In this antishadow scattering mode
 the elastic amplitude increases with decrease of the inelastic
 channels contribution.

The shadow scattering mode is often considered as the only possible
one. However, it should be noted that the two solutions of unitarity
 have an equal meaning and the antishadow scattering mode could
also be realized  at high energies in central collisions.
The both scattering modes are achieved in a continuous way in the
 $U$--matrix approach despite these modes are related with the two
 different solution of unitarity Eq. \ref{usol}.

Let us consider the transition to the antishadow scattering mode
 \cite{phl} in the framework of the $U$--matrix unitarization scheme. With
conventional parameterizations of the $U$--matrix in the form of Eq. \ref{eb2}
or Eq. \ref{eb1} the inelastic overlap function increases with energies
at modest values of $s$. It reaches maximum value $\eta(s,b)=1/4$ at some
energy $s=s_0$ and beyond this energy the transition to the antishadow
scattering mode occures. The region of small impact parameters corresponds
to this scattering mode when $Im f(s,b)> 1/2$ and
$\eta(s,b)< 1/4$. The quantitative analysis of the experimental data
 \cite{nuvc} has given prediction $\sqrt{s_0}=2$ TeV.

Thus, the function $\eta(s,b)$ becomes peripheral when energy is increasing.
At such energies the inelastic overlap function reaches its maximum
 value at $b=R(s)$ where $R(s)$ is the interaction radius.
So, beyond the transition threshold there are two regions in impact
 parameter space: the central region
of antishadow scattering at $b< R(s)$ and the peripheral region
of shadow scattering at $b> R(s)$.
At $b=R(s)$ the maximal absorbtion takes place (Fig. 1).

The transition to the antishadow scattering mode at small impact
parameters and high energies results also in a relatively slow growth
of inelastic cross--section:
\begin{equation}
\sigma_{inel}(s)=8\pi\int_0^\infty Im U(s,b)|1-iU(s,b)|^{-2}\sim\ln s.
\end{equation}
at $s\rightarrow \infty$.

It should be noted that  appearance of the antishadow scattering mode
does not contradict to the basic idea that the particle production
is the driving force for elastic scattering. Indeed, the imaginary
part of the generalized reaction matrix is the sum of inelastic channel
 contributions:
\begin{equation}
Im U(s,b)=\sum_n \bar{U}_n(s,b),
\end{equation}
where $n$ runs over all inelastic states and
\begin{equation}
\bar{U}_n(s,b)=\int d\Gamma_n |U_n(s,b,\{\xi_n\}|^2
\end{equation}
and $d\Gamma_n$ is the $n$--particle element of the phase space
volume.
The functions $U_n(s,b,\{\xi_n\})$ are determined by the dynamics
 of $2\rightarrow n$ processes. Thus, the quantity $ImU(s,b)$ itself
 is a shadow of the inelastic processes.
However, unitarity leads to self--damping of inelastic
channels \cite{bbl} and increase of the function $ImU(s,b)$ results in
decrease
 of the inelastic overlap function $\eta(s,b)$ when $ImU(s,b)$ exceeds unity.

At the energies when the antishadow mode starts to develop (it presumably
could already occur at energies of the Tevatron--Collider) the Pumplin
 bound Eq. \ref{pmb} for inelastic  diffraction dissociation
cannot be applied since
the main assumption used under its  derivation is no more valid.

The consideration of diffraction dissociation in the framework of
the $U$--matrix chiral quark model \cite{prd} shows that $\sigma_{diff}(s)$
 has a complicated energy dependence: it increases at energies when
 only the shadow scattering mode exists and decreases when the antishadow
scattering mode appears in the central hadron--hadron collisions.
Such a behavior of $\sigma_{diff}(s)$ reflects the changing energy dependence
of $\eta(s,0)$.

\section*{Conclusions}

Thus, loosely speaking the genesis of hadron scattering
 can be described as a transition from the grey to black disk and finally
to black ring with the antishadow scattering in the central region.
Such transformations are under control of unitarity
of the scattering matrix.
Of course, it would be interesting to consider  particular
physical origin of the antishadow scattering mode. First of all, the
existence of such  mode points out that  new phenomena
would be expected at high energies in the central hadronic collisions.
 Such collisions are usually associated with formation of quark--gluon
plasma and disoriented chiral condensate in the interior of interaction
region. What are the correlations between these phenomena and
the antishadow scattering mode? If there are any, it might be studied in
the framework of nonperturbative QCD  and in the
 experiments devoted to measuring observables in  soft processes
 (recent discussions of these problems are given in \cite{hal}).
 It seems  that the anomalies observed in cosmic ray experiments might
 also be correlated with  development of the  antishadow scattering mode
 in the  central hadron collisions.

\section*{Acknowledgements}
One of the authors (S. T.) would like to thank A. D. Krisch for
 warm hospitality at University of Michigan where this
 article was completed.

\newpage
\begin{figure}
\vspace{10cm}
\caption{The impact parameter dependence of inelastic overlap function
at energies $s>s_0$.}
\end{figure}
\end{document}